\documentstyle[amsfonts,amsmath,epsfig,12pt]{article}

\newcommand{\smallfrac}[2] {\mbox{$\frac{#1}{#2}$}}

\newcommand {\slsh} [1] {\not{\hbox{\kern-2pt${#1}$}}}

\def\drawbox#1#2{\hrule height#2pt
         \hbox{\vrule width#2pt height#1pt \kern#1pt
               \vrule width#2pt}
               \hrule height#2pt}

\def\Asym#1#2{\vcenter{\vbox{\drawbox{#1}{#2}
               \kern-#2pt       
               \drawbox{#1}{#2}}}}

\newcommand {\beq} {\begin{equation}}
\newcommand {\eeq} {\end{equation}}
  \newcommand {\ber}{\begin{eqnarray*}}
  \newcommand {\eer} {\end{eqnarray*}}
\newcommand {\bea}{\begin{eqnarray}}
  \newcommand {\eea} {\end{eqnarray}}

\newcommand{\Dslash}{\,{\raise.15ex\hbox{/}\mkern-12mu D}}

\begin{document}


\begin{titlepage}

\begin{center}
\vspace{1in}
\large{\bf Large-$N$ QCD and the Veneziano Amplitude}\\
\vspace{0.4in}
\large{Adi Armoni}\\
\small{\texttt{a.armoni@swansea.ac.uk}}\\
\vspace{0.2in}
\emph{Department of Physics, Swansea University}\\ 
\emph{Singleton Park, Swansea, SA2 8PP, UK}\\
\vspace{0.3in}
\end{center}

\abstract{We consider four scalar mesons scattering in large-$N_c$ QCD. Using the worldline formalism we show that the scattering amplitude can be written as a formal sum over Wilson loops. The AdS/CFT correspondence maps this sum into a sum over string worldsheets in a confining background. We then argue that for well separated mesons the sum is dominated by flat space configurations. Under additional assumptions about the dual string path integral we obtain the Veneziano amplitude. 
}
\end{titlepage}


The 1968 seminal paper by Veneziano marks the birth of string theory (``dual resonance model''). It concerns a proposal for the scattering amplitude of four scalar mesons, as follows \cite{Veneziano:1968yb}
\beq
{\cal A} (k_1, k_2, k_3,k_4) = { \Gamma (-\alpha(s)) \Gamma(-\alpha(t)) \over \Gamma (-\alpha (s) - \alpha(t))}  \label{veneziano1}
\eeq
with $\alpha (s)=\alpha(0) + \alpha ' s$, $\alpha (t)=\alpha(0) + \alpha ' t$ and $s,t$ are Mandelstam variables.

While the amplitude \eqref{veneziano1} is appealing due to its theoretical and phenomenological properties, it lacks a derivation from Quantum Chromodynamics (QCD). It is important to clarify the exact relation between the above amplitude and QCD. In particular, we wish to find whether there exists an approximation where we can derive the amplitude \eqref{veneziano1} from QCD. The topology of the string diagram (``disk amplitude'') suggests that it should be related to large-$N_c$ QCD with fixed $N_f$, namely to the 't Hooft limit of QCD where quark loops are suppressed (quarks are ``quenched''). As we shall see, the route from QCD to \eqref{veneziano1} involves additional approximations which are not controlled by QCD parameters. In particular, we assume that Wilson loops are calculated by flat space string worldsheets.

\begin{figure}[!ht]
\centerline{\includegraphics[width=4cm]{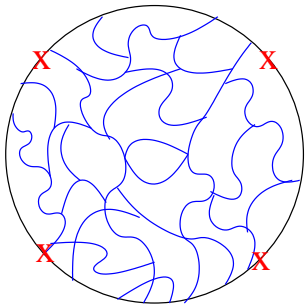}}
\caption{\footnotesize Perturbative description of the four mesons scattering amplitude as a ``fishnet'' diagram in planar QCD.}
\label{plot1}
\end{figure}

The first attempt to relate the Veneziano amplitude to field theory was made in 1970 \cite{Nielsen:1970bc,Sakita:1970ep}, even before the birth of QCD. Sakita and Virasoro argued \cite{Sakita:1970ep}, using a scalar field theory as a model of strong interactions, that the field theory amplitude should look like a dense fishnet, see fig.\eqref{plot1}. Moreover, they argued that at high orders in perturbation theory the gaps between the holes in the fishnet close and the amplitude should resemble a string amplitude, namely the value of the amplitude should correspond to the area of the string worldsheet. While this idea can easily be implemented in large-$N_c$ QCD (with a fishnet made of gluons),  it cannot be the full story as it lacks the essential non-perturbative effects which are responsible for confinement with a string tension and a mass gap. It is therefore necessary to address the problem by using a formalism that incorporates non-perturbative effects.

We consider $SU(N_c)$ QCD with $N_f$ massless quarks in the 't Hooft limit, where $g^2 N_c$ and $N_f$ are kept fixed. The partition function takes the form
\beq 
{\cal Z}=\int DA_\mu \exp (-S_{\rm YM}) \left (\det (i\slsh D) \right )^{N_f}
\eeq
Let us write the fermionic determinant as a sum over (super-)Wilson loops by using the worldline formalism \cite{Strassler:1992zr}
\beq
 \left (\det (i\slsh D) \right ) ^{N_f} = \exp( N_f \Gamma [A_\mu]) \, ,
\eeq
with
\bea
\label{wlineint}
 \Gamma [A_\mu] &=&
-{1\over 2} \int _0 ^\infty {dT \over T}
\nonumber\\[3mm]
 &\times&
\int {\cal D} x^\mu {\cal D}\psi ^\mu
\, \exp
\left\{ -\int _{\epsilon} ^T d\tau \, \left ( {1\over 2} \dot x ^\mu \dot x ^\mu + {1\over
2} \psi ^\mu \dot \psi ^\mu \right )\right\}
\nonumber \\[3mm]
 &\times &  {\rm Tr }\,
{\cal P}\exp \left\{   i\int _0 ^T d\tau
\,  \left (A_\mu \dot x^\mu -\frac{1}{2} \psi ^\mu F_{\mu \nu}  \psi ^\nu
\right ) \right\}  \, ,
\eea
where $\mu,\nu=0,...,3$.

The expansion of the exponent in powers of $N_f$ 
\beq
\exp( N_f \Gamma )\sim 1+ N_f \Gamma + {\cal O}(N_f^2)
\eeq
is in fact an expansion in $N_f/N_c$, since $\langle W_1 W_2 ... W_l \rangle _{\rm conn.} \sim N_c^2 (N_f/N_c)^l$. Therefore, as we shall see, in the 't Hooft limit the dominant class of diagrams for meson scattering involves one Wilson loop\cite{Armoni:2008jy}.

Let us consider the scattering amplitude of four scalar mesons
\beq
{\cal A} (x_1, x_2, x_3,x_4)= \langle \bar qq (x_1) \bar qq (x_2) \bar qq(x_3) \bar qq (x_4) \rangle \,.
\eeq
For simplicity we assume flavor-singlet mesons. Flavored mesons can be easily incorporated by adding ``Chan-Paton'' factors, resulting in an overall group theoretic factor of ${\rm Tr}\, (\lambda ^1 \lambda ^2 \lambda ^3 \lambda ^4)$. In order to calculate the amplitude we add to the action a source of the form $\int d^4 x \, J(x) \bar qq$
\beq
 {\cal Z}_J =\int DA_\mu \exp (-S_{\rm YM}) \left (\det (i\slsh D + J(x) )\right )^{N_f} 
\eeq
 and differentiate the partition function with respect to $J(x_1),J(x_2),J(x_3),J(x_4)$
\beq
 \langle \bar qq (x_1) \bar qq (x_2) \bar qq(x_3) \bar qq (x_4) \rangle \rangle 
= {\delta \over \delta J(x_1)} {\delta \over \delta J(x_2)} {\delta \over \delta J(x_3)} {\delta \over \delta J(x_4)} {\log \cal Z}_J |_{J=0}
\eeq

In the worldline formalism, such a variation excludes Wilson loops whose contours do not pass through $\{ x_1,x_2,x_3,x_4\}$ \cite{Makeenko:2009rf}\cite{Armoni:2008jy}.

\begin{figure}[!ht]
\centerline{\includegraphics[width=4cm]{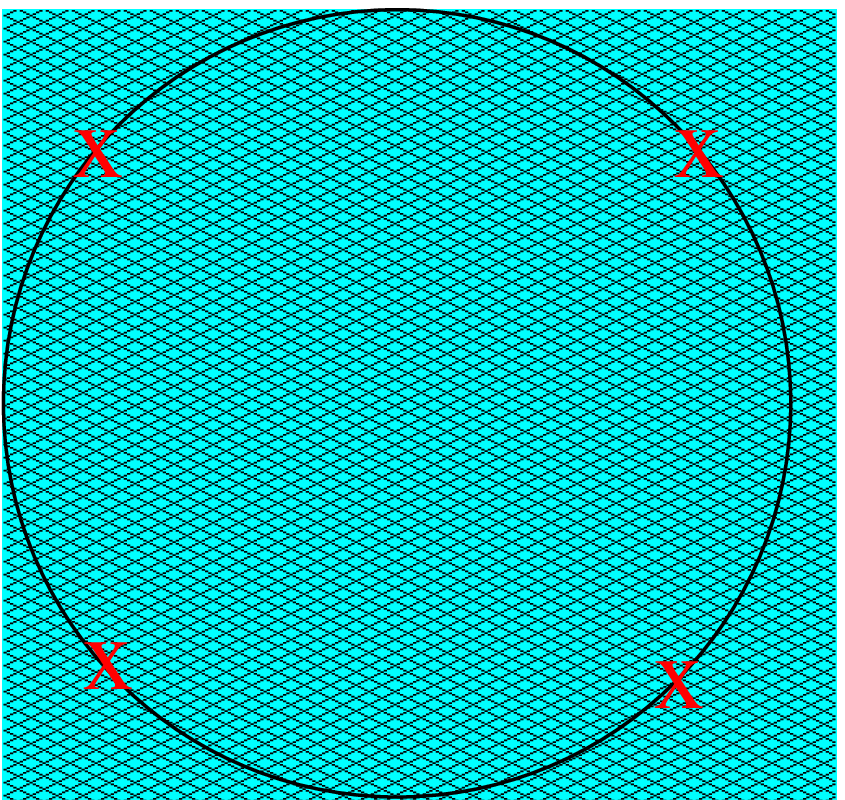}}
\caption{\footnotesize Worldline description of the four mesons scattering amplitude. It resembles the string disk amplitude.}
\label{plot2}
\end{figure}

Thus in the 't Hooft limit the leading large-$N_c$ expression for the scattering amplitude can schematically be written as
\beq
{\cal A} (x_1, x_2, x_3,x_4)=
 {1\over 2} \int _0 ^\infty {dT \over T}\int {\cal D} x
\, \exp
\left ( -\int  d\tau \, \smallfrac{1}{2} \dot x^2 _\mu  \right )\langle W(x_1,x_2,x_3,x_4) \rangle _{\rm YM} \,.
\eeq
In order to simplify the notation we omitted the worldline fermions in the above expression. The Wilson loops are calculated in the large-$N$ pure Yang-Mills theory. The sum is over all sizes and shapes of Wilson loops that pass through the points $\{ x_1,x_2,x_3,x_4\}$. The QCD amplitude admits the topology of a disk and it resembles the string disk amplitude \eqref{plot2} \cite{Makeenko:2012va}.

In order to proceed we use holography to evaluate $\langle W(x_1,x_2,x_3,x_4) \rangle _{\rm YM} $ (instead we could have used the lattice strong coupling expansion). The holographic prescription for evaluating the Wilson loop expectation value is to find the minimal worldsheet that ends on the boundary of the space (anti de Sitter in the case of ${\cal N}=4$ super Yang-Mills theory) and passes through $\{ x_1,x_2,x_3,x_4\}$ \cite{Maldacena:1998im,Rey:1998ik}.

We therefore {\it propose} that the field theory path integral over all sizes and shapes of contours that pass through  $\{ x_1,x_2,x_3,x_4\}$, translates into the Polyakov path integral for string worldsheets with a disk topology that end on the boundary of the space and pass through $\{x_1,x_2,x_3,x_4\}$
\bea
& & {1\over 2} \int _0 ^\infty {dT \over T}\int {\cal D} x^\mu(\tau)
\, \exp
\left ( -\int  d\tau \, \smallfrac{1}{2} \dot x^2 _\mu  \right )\langle W(x_1,x_2,x_3,x_4) \rangle = \label{holography} \\
& & \int {\cal D} g^{\alpha \beta}\, {\cal D} x^M
\, \exp
\left ( -{1\over 2\pi \alpha '}\int   d^2\sigma \, \sqrt g g^{\alpha \beta} \partial _\alpha x^M  \partial _\beta x^N G_{MN} \right )|_{\{x_1,x_2,x_3,x_4\}} \nonumber
\eea
with $M,N=0,...,D$ ($D=9$ for the superstring). The above proposal \eqref{holography} is conjectured to hold for all gauge/gravity dual pairs. The information about the specific gauge theory is encoded in the metric $ G_{MN}$. We will use Witten's model of back-reacted $N_c$ D4 branes compactified on a thermal circle \cite{Witten:1998zw}. The conjecture is \cite{Witten:1998zw} that type IIA string theory on
\bea
& & ds^2 =(U/R)^{3/2}(\eta _{\mu \nu} dx^\mu dx^\nu +f(U)d\tau ^2)+ (R/U)^{3/2} ({dU^2 \over f(U)} + U^2 d\Omega^2 _4) \nonumber \\
& & f(U)= 1-U^3_{KK}/U^3  \, ,
\eea
with anti-periodic condition for NS-R modes on $\tau$ is dual to large-$N$ Yang-Mills theory. Unfortunately, in this model the scale of the Kaluza-Klein modes and the scale of confinement are of the same order, $\Sigma \sim (g^2 N_c) M^2_{KK}$ \cite{Kruczenski:2003uq}. Note that since we use the type IIA superstring, we should use in \eqref{holography} the Polyakov action for the superstring, namely we should incorporate worldsheet fermions. 

A calculation of Wilson loop expectation value in Witten's background yields an area law for large loops, $\langle W \rangle = \exp (-\Sigma A)$, with a string tension $\Sigma={1\over 2\pi \alpha '} ({U_{KK} \over R})^{3\over 2}$. The reason is that for a sufficiently large loop the string will quickly drop to $U=U_{KK}$ and the main contribution to the action is essentially by a flat space worldsheet that resides close to $U_{KK}$ \cite{Brandhuber:1998er}.

\begin{figure}[!ht]
\centerline{\includegraphics[width=6cm]{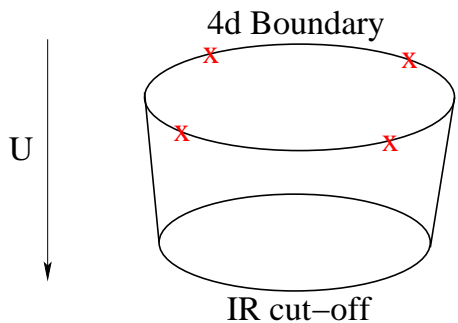}}
\caption{\footnotesize A string worldsheet in a confining background. The string ``sits'' at the IR cut-off. }
\label{plot3}
\end{figure}

It is therefore anticipated that if the points $\{x_1,x_2,x_3,x_4\}$ are well separated from each other $|x_i-x_j| \gg 1/ \sqrt \Sigma$, the sum over string worldsheets will be dominated by flat space configurations, as depicted in fig.\eqref{plot3}. This claim is supported by the analysis of ref.\cite{Makeenko:2009rf}, where the authors showed that under the assumption of confinement (area law) the sum over Wilson loops is dominated by a saddle-point that corresponds to the IR piece of the worldsheet in fig.\eqref{plot3}. Our claim is indeed similar to that of ref.\cite{Makeenko:2009rf}: assuming that Wilson loop expectation values are calculated by a flat space string configuration implies that they admit an area law.

String worldsheets that span in the compact directions ($\tau$ and the four-sphere) are expected to admit a larger action than string worldsheets that reside in the flat four dimensional space. 

If this is indeed the case, we proceed as follows. We first fix the gauge by choosing a flat worldsheet metric and add ghosts. At tree level the ghosts decouple from $x^\mu$ and therefore will not play a role in the calculation of the disk amplitude, so they will be omitted. Since we anticipate that the contribution of compact directions ($\tau$, $\Omega$ and their superpartners) is not dominant, we assume that we can carry out the integration over these fields as if the metric is flat. In fact, if the metric was $U$ independent, the integration over the compact directions becomes indeed trivial. 

 Under these assumptions the calculation is crudely approximated by the path integral  
\beq 
\int {\cal D} x^\mu {\cal D} U
\, \exp
\left ( -{1\over 2\pi \alpha '}\int d^2\sigma \,\{ ( \partial _\alpha x^\mu  \partial ^\alpha x^\nu G_{\mu \nu}) + ( \partial _\alpha U  \partial ^\alpha U G_{UU}) \}  \right )|_{\{x_1,x_2,x_3,x_4\}} \, , \label{polyakov}
\eeq
with the 5d metric
\beq
 ds^2 =(U/R)^{3/2}\eta _{\mu \nu} dx^\mu dx^\nu +  (R/U)^{3/2} ({dU^2 \over f(U)})\, .
\eeq
Due to the metric singularity, the path integral \eqref{polyakov} is dominated by $U$ in the vicinity of $U_{KK}$. This is the case for large worldsheets, where the string ``sits'' at the horizon. Thus, let us insert $\delta (U-(U_{KK}+\epsilon ))$ into the path integral, suppressing quantum fluctuations in the $U$ directions. Later on we will discuss what is expected to happen if we omit the delta function and allow fluctuations in the $U$ direction. The modified path integral reads
\bea
& &
\int {\cal D} x^\mu  {\cal D} U \delta (U-U_{KK}-\epsilon) \times \nonumber \\
& & \times \exp
\left ( -{1\over 2\pi \alpha '}\int d^2 \sigma \, \{ ( \partial _\alpha x^\mu  \partial ^\alpha x^\nu G_{\mu \nu}) + ( \partial _\alpha U  \partial ^\alpha U G_{UU}) \}  \right )|_{\{x_1,x_2,x_3,x_4\}}  \nonumber \\  
& &
= \int {\cal D} x^\mu \, \exp \left ( -{1\over 2\pi \alpha '}\int d^2 \sigma \, \partial _\alpha x^\mu  \partial ^\alpha x^\nu \hat G_{\mu \nu} \right )|_{\{x_1,x_2,x_3,x_4\}} \, , \label{polyakov2}
\eea
with $\hat G_{\mu \nu}$ the flat 4d metric
\beq
 ds^2 =(U_{KK}/R)^{3/2}\eta _{\mu \nu} dx^\mu dx^\nu \, .
\eeq

It is easier to calculate the amplitude in momenta space. Let us carry out the computation with external states of $k_i^2=0$. This can be justified by assuming a scattering of four massless scalars. A physical case that corresponds to massless (pseudo-)scalars is the scattering of pions - the Nambu-Goldstone bosons that correspond to chiral symmetry breaking.

To this end we assume that the appropriate scalar vertex operator to insert into the path integral takes the form
\beq
 V \sim \int dy \, \exp (ik\cdot x) \label{vertex}
\eeq
 (the vertex operators sit on the disk boundary). Note that due to our assumption that $U=U_{KK}$ we can use flat space vertex operators. Since our bulk theory is type IIA the vertex operators involved in scalar scattering are anticipated to be supersymmetric and not the assertion \eqref{vertex}. The proposal \eqref{vertex} is motivated by the fact that supersymmetry is broken. Since the {\it only} massless states in the gauge theory are the massless Nambu-Goldstone bosons (the ``pions''), the vertex operator \eqref{vertex} of the dual string must correspond to these states, hence it is suggested to be a purely bosonic vertex operator. The assertion \eqref{vertex} is the simplest operator which involves massless bosonic fields.

Inserting vertex operators in the path integral is not equivalent to a simple Fourier transform, but it rather leads to an element of the scattering matrix, see \cite{Pius:2014gza} for a recent discussion.  
\bea
&& {\cal A}(k_1,k_2,k_3,k_4)= \\
& & \int {\cal D} x^\mu \prod _{i=1,..,4} \int dy _i \exp \left (ik_i x(y_i) \right ) \, \exp
\left ( -\Sigma \int d^2 \sigma \, \partial _\alpha x^\mu  \partial ^\alpha x _\mu  \right ) \, , \nonumber
\eea
with $y_i$ a coordinate on the disk boundary. The scattering amplitude is
\beq
 {\cal A}(k_1,k_2,k_3,k_4) \sim \delta (\sum _i k_i) \int \prod _i dy _i \prod _{j<l} (y_j-y_l)^{2\alpha ' (R/U_{KK})^{3/2} k_j \cdot k_l} \, ,
\eeq
with $i,j,l=1..4$. Since we started with type IIA string theory and we carried out a calculation in flat space, the result is the celebrated Veneziano amplitude for the superstring with a string tension $\Sigma$.

 An important issue is the suppression of fluctuations in the $U$ direction. Although classically the string sits at $U=U_{KK}$, quantum mechanically it fluctuates. Incorporating such fluctuations is not an easy task, since it involves a solution of the non-linear sigma model. The fluctuations near $U_{KK}$ are expected to affect the low lying spectrum of the string and to lead to deviations from the exact linearity of the Regge trajectories that characterise the Veneziano amplitude.

In conclusion, we have shown that the large-$N_c$, fixed $N_f$, QCD disk amplitude that represents scalar mesons scattering can be written as a sum over Wilson loops. By using holography the calculation is mapped into a sum over string worldsheets. Carrying out this sum for well separated mesons (or small $\Delta k_i$, corresponding to the low-energy regime) ignoring fluctuations in the $U$ direction, leads to a flat space calculation which yields the Veneziano amplitude for the type IIA superstring.

Note that while we used holography to relate Wilson loops in the gauge theory side to string worldsheets, we did not need to use the intuitive relation between mesons and open strings. Our approach is different from other approaches that relate gauge theory scattering amplitudes with string amplitudes. In particular, in contrast to Polchinski and Strassler \cite{Polchinski:2001tt}, we are interested in meson scattering and not in glueball (pomeron) scattering and, moreover, we use the method of \cite{Armoni:2008jy} in order to introduce mesons as opposed to the more common method of using a probe brane.
  
We therefore demonstrated how string disk amplitudes are related to QCD amplitudes. Our analysis can be easily generalised to the case of $k$-mesons scattering. A more challenging task is to take into account the spacetime curvature and fluctuations in the $U$ directions in order to obtain corrections to the Veneziano amplitude. If there were a gravity dual to large-$N$ QCD and if we could carry out the full path integral we should have obtained an amplitude that describes correctly the DIS regime of QCD. Finally, it is expected that some of the perimeter effects will account for corrections due to quark masses, as in \cite{Kruczenski:2004me}. 

\vskip 1cm
{\it \bf Acknowledgments.} It is a pleasure to thank Edwin Ireson, Prem Kumar, Carlos Nunez, Maurizio Piai and Gabriele Veneziano for fruitful discussions.


\end{document}